\documentclass[10pt,twocolumn]{revtex4}
\usepackage{slashed,graphicx}
\tighten

\begin{document}
%\draft
\title{Neutron star properties and the equation of state   \\
of neutron-rich matter}
\author{P. G. Krastev and F. Sammarruca}
\address{Physics Department, University of Idaho, Moscow, ID 83844, U.S.A}
\date{\today}

\email[P. Krastev: ]{kras7125@uidaho.edu}

\email[F. Sammarruca: ]{fsammarr@uidaho.edu}

\begin{abstract}
We calculate total masses and radii of neutron stars (NS) for pure
neutron matter and nuclear matter in $\beta$-equilibrium. We apply a
relativistic nuclear matter equation of state (EOS) derived from
Dirac-Brueckner-Hartree-Fock (DBHF) calculations. We use realistic
nucleon-nucleon (NN) interactions defined in the framework of the
meson exchange potential models. Our results are compared with other
theoretical predictions and recent observational data. Suggestions
for further study are discussed.
\end{abstract}
\maketitle

%\begin{multicols}{2}
\renewcommand{\thesection}{\arabic{section}}
\section{Introduction}
One of the most important and challenging problems in both
theoretical and experimental nuclear physics is to understand the
properties of matter under extreme conditions of density and
pressure. The determination of the EOS (namely, the relationship
between pressure and density) associated with such matter is a
non-trivial problem which has attracted significant effort over the
last few decades. Concerning terrestrial systems, detailed knowledge
of the EOS is important, for instance, for understanding heavy-ion
collision dynamics. On the other hand, the EOS is crucial for
determining the properties of one of the most exotic objects in the
universe--neutron stars. Neutron star properties depend on the
knowledge of the EOS over a wide range of densities -- from the
density of iron at the stellar surface up to several times the
density of normal nuclear matter in the core region of the star
\cite{Weber99}.

Neutron stars are the smallest and densest stars known to exist.
Like all stars neutron stars rotate--some at a rate of a few hundred
revolutions per second. Such a fast rotating object will experience
enormous centrifugal  force which must be balanced by another force,
gravity in this case, to prevent the star from falling apart. The
balance of the two forces sets the lower limit on the stellar
density \cite{Glendenning}. Some neutron stars are in binary orbit
with a companion and, in some cases, the application of orbital
dynamics allows for an assessment of the masses. So far six neutron
star binaries are known and all of them have masses in the range
$1.36\pm0.08M_{\odot}$ \cite{Thorsett&Chakrabarty}. Clearly,
observations of NS masses and radii impose important constraints on
the EOS of dense matter, as the latter constitutes the basic input
quantity that enters the structure equations of a neutron star
\cite{Weber99,Wang,Shapiro83}.

Although considerable progress has been made, the EOS of dense
matter remains uncertain at densities above $\rho_n\simeq
3\times10^{14}g/cm^{3}$ \cite{Ostgaard}. For the reason stated
above, this places uncertainty on the calculated NS properties.
 Determining the high density
stellar matter EOS is a tremendous task. While at densities
$\rho\simeq\rho_0$, ($\rho_0\simeq 0.17 fm^{-3}$ is the normal
nuclear density), matter consists mainly of nucleons and leptons, at
higher densities several other species of particles are expected to
appear due to the rapid rise of the baryon chemical potentials with
density \cite{Baldo&Burgio2000}. Among these are strange baryons,
such as the $\Lambda$, $\Sigma$, and $\Xi$ hyperons. In addition,
pion-nucleon resonances may appear in stellar matter along with pion
and kaon condensations. Moreover, at super high densities nuclear
matter is expected to undergo a transition to quark-gluon plasma.
The  value of the transition density may be obtained from QCD
lattice calculations at finite baryon densities, but is presently
still uncertain \cite{Baldo&Burgio2000}.

After the first theoretical calculations of neutron star properties
performed by Oppenheimer and Volkoff \cite{Oppenheimer&Volkoff39}
and independently by Tolman \cite{Tolman39}, many theoretical
predictions appeared in literature. Non-relativistic and
relativistic approaches have been used. A number of early
theoretical investigations on NS properties were done within the
non-relativistic Skyrme framework. The reader is referred to the
work of F. Douchin and P. Haensel in Ref.~\cite{Douchin&Haensel01}.
The conventional Brueckner theory with a continuous choice for the
single particle potential and three-body forces was applied in
Ref.~\cite{Baldo&Burgio98}. Numerous predictions allow for hyperons
together with nucleons and leptons in generalized
$\beta$-equilibrium to be included in the stellar interior. The
Brueckner-Hartree-Fock (BHF) scheme was extended to include these
contributions \cite{Baldo&Burgio&Schulze98,Schulze98} and applied in
NS calculations \cite{Baldo&Burgio2000}. Since the Walecka model
\cite{Walecka86} was proposed and applied to nuclear matter
properties, the relativistic mean field approach has been widely
used in determining NS total masses and radii
\cite{Li2004,Lawley2004}. The DBHF approach was used to compute NS
properties in Refs.~\cite{Ostgaard,Sumiyoshi95}.

In general, the EOS of stellar matter is considerably model
dependent and, as a result, predictions of NS properties are quite
different from model to model. Among the sources of model dependence
are \cite{Weber99}: (1) the many-body framework used for the
determination of the EOS, (2) the model used for the bare NN
interaction, (3) the kinds of hadrons/leptons included in the
description of electrically neutral NS matter, (4) the treatment of
pion and kaon condensations, (5) including the effects of fast
rotation, (6) allowing for a phase transition from confined hadronic
matter to deconfined quark matter. The possibility of such
transition has attracted great interest over the last 30 years
\cite{QM1,QM2,QM3,QM4}. It is generally agreed upon that the high
pressure found in the core of a neutron star creates an ideal
physical environment for hadrons to transform into quark matter -- a
state of matter of practically infinite lifetime. Nevertheless,
until recently no empirical evidence had been reported. To a certain
extent, this may signify that the occurrence (or not) of a
transition to quark matter in the NS interior has only a minor
impact on the overall static properties (range of possible masses,
radii, limiting rotational periods) of the star \cite{Weber99}. The
situation, however, could be completely different for the timing
structure of a pulsar (its braking behavior as it evolves) which is
expected to diverge significantly from the normal pulsar behavior if
the star converts a fraction of its core matter to pure quark matter
\cite{Weber99}.

Recently we have been concerned with probing the behavior of the
isospin-asymmetric EOS \cite{Alonso&Sammarruca1}. In our work on
neutron skins \cite{Alonso&Sammarruca2}, we have studied
applications of the EOS at densities typical for normal nuclei. In
Ref.~\cite{SBK} we discussed the one-body potentials for protons and
neutrons from DBHF calculations of neutron rich matter (in
particular their dependence upon the degree of asymmetry in proton
and neutron concentrations). More recently, we applied our
microscopic approach in calculations of effective in-medium NN cross
sections \cite{SK}. It is also important and timely to look into
systems which are likely to constrain the behavior of the EOS at
higher densities, where the largest model dependence is observed.
This is the main purpose of the present paper, where we report
predictions for masses and radii of static (non-rotating) neutron
stars. Aside from minor adjustments, the framework is the one
described in Ref.~\cite{Alonso&Sammarruca1}. We apply a relativistic
(DBHF-based) EOS and perform our calculations for two distinct
cases: (1) pure neutron matter, and (2) baryon/lepton matter in
$\beta$-equilibrium. It is also our objective to provide an overview
of the present status of both theory and experimental constraints.
Together with systematic calculations based on our microscopic
model, this broad outlook will help us gauge the quality of our
tools and determine the importance of potentially missing
mechanisms.

This work is organized in the following way: after the introductory
notes in this section,  we briefly review our theoretical framework
(section 2); the relativistic (DBHF-based) EOS for both neutron
matter and $\beta$-stable matter is discussed in section 3; our
numerical results are presented and discussed in section 4; we
conclude in section 5 with a short summary and suggestions for
further studies.

\section{Formalism}
In order to provide the reader with a self-contained manuscript, in
this section we outline briefly the formalism used to derive the
isospin-asymmetric EOS. A more extensive discussion can be found in
Ref.~\cite{Alonso&Sammarruca1} and references therein.

\renewcommand{\thesubsection}{2.\arabic{subsection}}
\subsection{Realistic nucleon-nucleon interactions}

The starting point of any microscopic calculation of nuclear
structure or reactions is a realistic free-space NN interaction. A
realistic and quantitative model for the nuclear force that has a
reasonable basis in theory is the one-boson-exchange (OBE) model
\cite{Machleidt89}. In this framework bosons with masses below the
nucleon mass are typically included. The model we apply in the
present study consists of six bosons four of which are of major
importance: (1) The pseudoscalar pion with a mass of about 138 MeV.
It is the lightest meson and provides the long-range part of the NN
potential and most of the tensor force; (2) The $\rho$ vector meson
-- a $2\pi$ P-wave resonance with a mass of about 770 MeV and spin
1. Its major role is to cut down the pion tensor force at short
range; (3) The $\omega$ vector meson -- a $3\pi$ resonance of about
783 MeV and spin 1. It creates a strong repulsive central force of
short range and the short-ranged spin-orbit force; (4) The
isoscalar-scalar $\sigma$ boson with a mass of about 550 MeV. It
provides the intermediate range attraction necessary for nuclear
binding and can be understood as a simulation of the correlated
S-wave $2\pi$-exchange.

The mesons are coupled to the nucleon through the following
meson-nucleon Lagrangians for the pseudoscalar (ps), scalar (s), and
vector ($\upsilon$) fields, respectively:

\begin{equation}
{\cal{L}}_{p\upsilon}=-\frac{f_{ps}}{m_{ps}}\bar{\psi}\gamma^5\gamma^{\mu}
\psi\partial_{\mu}\phi^{(ps)}
\end{equation}
\begin{equation}
{\cal{L}}_s=g_s\bar{\psi}\psi\phi^{(s)}
\end{equation}
\begin{equation}
{\cal{L}}_{\upsilon}=
-g_{\upsilon}\bar{\psi}\gamma^{\mu}\phi_{\mu}^{(\upsilon)}
-\frac{f_{\upsilon}}{4m}\bar{\psi}\sigma^{\mu\nu}\psi
(\partial_{\mu}\phi^{(\upsilon)}_{\nu}-\partial_{\nu}\phi^{(\upsilon)}_{\mu})
\end{equation}
with $\psi$ the nucleon and $\phi_{\mu}^{(\alpha)}$ the meson fields
respectively (notation and conventions as in
Ref.~\cite{Bjorken&Drell}). For isovector (isospin 1) mesons (such
as $\rho$ and $\pi$), $\psi^{(\alpha)}$ is to be replaced by
$\tau\cdot\phi^{(\alpha)}$, with $\tau$ the usual Pauli matrices.

From the above Lagrangians the OBE amplitudes can be derived (see,
for instance, Ref.~\cite{Machleidt89}). The OBE potential is defined
as the sum of the OBE amplitudes of all exchanged mesons.

In this work, we use the Bonn A, B, and C parameterizations of the
OBE potential formulated in the framework of the Thompson equation.
The main difference between these three potentials is in the
strength of the tensor force as reflected in the predicted D-state
probability of the deuteron, $P_D$ \cite{Brockmann&Machleidt96}.
Bonn A has the weakest tensor force with $P_D=4.5\%$ (see Table 2 in
Ref.~\cite{Brockmann&Machleidt96}). The Bonn B and C potentials
predict 5.1\% and 5.5\%, respectively. It is well known that the
strength of the tensor force is a crucial factor in determining the
location of the nuclear matter saturation point on the Coester band
\cite{Coester}.

\subsection{Conventional Brueckner Theory}

We review here the main steps leading to the self-consistent
calculation of the energy per particle in infinite nuclear matter,
with or without isospin asymmetry, within the BHF approach. This is
mainly to provide a baseline for the next subsection where we
describe its relativistic extension (which is the method we actually
apply). For a complete description of the BHF method the reader is
referred to Refs.~\cite{Bethe71,Haftel&Tabakin70,Sprung72}. For a
review, see, for example, Ref.~\cite{Machleidt89}.

Nuclear matter is characterized by its total density or Fermi
momentum $k_F$ (and the degree of isospin asymmetry, in case of
unequal proton and neutron densities). With $k_1$ and $k_2$ being
the momenta of two nucleons with respect to nuclear matter, it is
customary to formulate the problem in terms of their relative
momentum, ${\bf{K}}=\frac{1}{2}({\bf{k}}_1-{\bf{k}}_2)$, and
one-half of the center-of-mass momentum,
${\bf{P}}=\frac{1}{2}({\bf{k}}_1+{\bf{k}}_2)$. (Clearly,
${\bf{k}}_{1,2}={\bf{P}}\pm{\bf{K}}$.)

The effective nucleon-nucleon interaction in nuclear matter is
described in terms of the reaction matrix, or the Brueckner
$G$-matrix, which satisfies the in-medium scattering equation (the
Bethe-Goldstone equation)

\begin{eqnarray}
G_{ij}({\bf{P;K,K_0}})&=&V_{ij}({\bf{K,K_0}})-\int\frac{d^3K'}{(2\pi)^3}
V_{ij}({\bf{K,K'}}) \nonumber\\
&&\times\frac{Q_{ij}({\bf{K',P}};k_F)G_{ij}({\bf{P;K',K_0}})}
{\epsilon_{ij}^*({\bf{P,K'}})-(\epsilon_{ij}^*)_0({\bf{P,K_0}})}
\end{eqnarray}
where $ij$=$nn$, $pp$, or $np$, and the asterisk signifies that
medium effects are applied to these quantities. $\bf{K_0}$,
$\bf{K}$, and $\bf{K'}$ are the initial, final, and the intermediate
momenta, respectively. $V_{ij}$ is the two body OBE potential
briefly discussed in the previous subsection. $\epsilon_{ij}^*$ is
the energy of the two-nucleon system, and $(\epsilon_{ij}^*)_0$ is
the starting energy. Thus,
\begin{equation}
\epsilon_{ij}^*({\bf{P,K}})=e_i^*({\bf{P,K}})+e_j^*({\bf{P,K}}),
\end{equation}
with $e_{i,j}^*$ the total energy of a single nucleon in nuclear
matter.

The $G$-matrix equation (Eq.~(4)) is density dependent due to the
presence of the Pauli projection operator, Q, which prevents
scattering into occupied states and is defined as
\begin{equation}
Q_{ij}({\mathbf{K,P}},k_F)=\left\{
\begin{array}{l l}
1 & \quad \mbox{if $k_1>k_F^i$ and  $k_2>k_F^j$}\\
0 & \quad \mbox{otherwise.}
\end{array}
\right.
\end{equation}
Detailed expressions for the (angle-averaged) Pauli operator can be
found, for instance, in Ref.~\cite{Alonso&Sammarruca1} for the case
of unequal proton and neutron Fermi momenta.

Equation (4) is density dependent also through the single-particle
energy,
\begin{equation}
e_i^*=T_i(p)+U_i(p),
\end{equation}
with $T_i(p)$ the kinetic energy and $U_i(p)$ the potential energy
due to the interaction of the nucleon with all the others in the
medium. We define
\begin{eqnarray}
U_i(p)=<p|U_i|p> & = & Re[\sum_{q\leq k_F^n}<pq|G_{in}|pq-qp>\nonumber\\
&&+\sum_{p\leq k_F^p}<pq|G_{ip}|pq-qp>]
\end{eqnarray}
with $|p>$ and $|q>$ single particle momentum, spin, and isospin
states. Schematically, Eq.~(8) has the form
\begin{equation}
U_n=U_{nn}+U_{np}
\end{equation}
\begin{equation}
U_p=U_{pp}+U_{pn},
\end{equation}
for neutrons and protons, respectively. Together with the $G$-matrix
equation, Eqs.~(9-10) constitute a coupled, self-consistency
problem, which is solved using the ``effective mass approximation''
\cite{Haftel&Tabakin70}.

To demonstrate the self-consistency procedure, we take the
non-relativistic single-particle energy \cite{Haftel&Tabakin70} and
set
\begin{equation}
\frac{p_i^2}{2m_i}+U_i(p_i)=\frac{p_i^2}{2m_i^*}+U_{0,i},
\end{equation}
which amounts to parametrizing the potential $U_i(p)$ in terms of
the nucleon effective mass $m_i^*$ and the constant $U_{0,i}$
($i=n,p$). Equation (11) implies
\begin{equation}
U_i(p)=\frac{1}{2}\frac{m_i-m_i^*}{m_im_i^*}p^2+U_{0,i}.
\end{equation}
Hence in the non-relativistic case the single-particle potential,
$U_i(p)$, has been fitted with a quadratic function of $p$. (The
subscript ``$i$'' signifies that the parameters are different for
neutrons and protons.) Starting with some initial values of $m_i^*$
and $U_{0,i}$, the $G$-matrix equation is solved and a first
solution for $U_i(p)$ is obtained. These solutions are then
parameterized in terms of a new set of constants, and the procedure
is repeated until convergence is reached (that is, until differences
between the parameters from successive iterations are within the
desired accuracy). Once the single-particle potential is available,
we calculate the (average) energy per particle (to lowest order in
the $G$-matrix).

As is well known, nuclear matter calculations based on the BHF
approach fail to predict correctly the saturation properties of
nuclear matter. The typical trend is that the saturation density is
too high for reasonable energies. Different calculations based on
the conventional BHF approach differ somewhat in their predictions
of nuclear matter saturation properties, but all of them fail to
predict simultaneously the correct saturation energy and
density~\cite{Machleidt89}. For this reason, it has become popular
to implement non-relativistic calculations with contributions from
phenomenological three-body forces.

On the other hand, already more than two decades ago it was realized
that the explicit treatment of the lower component of the Dirac
spinor in the medium could provide the missing saturation mechanism.
This observation started what became known as the DBHF approach. We
review some of the main points in the next section.

\subsection{The Dirac-Brueckner-Hartree-Fock approach}

For a detailed description see, for example,
Refs.~\cite{Brockmann&Machleidt90,Horowitz&Serot84,Haar&Malfleit87}.
The essential point of the DBHF approach is to use the Dirac
equation for the single particle motion in nuclear matter
\begin{equation}
(\slashed{p}_i-m_i-U_i(p))u_i({\bf{p}},s)=0,
\end{equation}
where the most general Lorentz structure for
 $U_i(p)$ is approximated as \cite{Brockmann&Machleidt90}
\begin{equation}
U_i(p)\approx U_{S,i}(p)+\gamma_{0}U^0_{V,i}(p).
\end{equation}
Here $U_{S,i}$ is an attractive scalar field and $U^0_{V,i}$ is the
timelike component of a repulsive vector field. The fields $U_{S,i}$
and $U^0_{V,i}$ are in the order of several hundred MeV and strongly
density dependent \cite{Brockmann2000}. With the definitions
$m_i^*(p)=m_i+U_{S,i}(p)$ and $(p_i^0)^*=p_i^0-U^0_{V,i}(p)$, the
Dirac equation in nuclear matter can be written as
\begin{equation}
(\slashed{p}_i^*-m_i^*)u_i({\bf{p}},s)=0.
\end{equation}
The positive energy solution of Eq.~(15) is given by
\begin{equation}
u_i({\bf{p}},s)=\left(\frac{E_i^*(p)+m_i^*}{2m_i^*}\right)^{1/2}\left(
\begin{array}{c}
1\\
\frac{\sigma\cdot{\bf{p}}}{E_{p,i}^*+m_i^*}
\end{array}
\right)\chi_S,
\end{equation}
with $\chi_S$ a Pauli spinor, and
$E_i^*(p)=(m_i^*{^2}+{\bf{p}}^2)^{1/2}$. We notice that this is
formally identical to the free-space spinor, but with $m_i$ replaced
by $m_i^*$.

The single particle potential is then calculated from the operator
given in Eq.~(14) and properly normalized Dirac spinors. Namely,
\begin{equation}
U_i(p)=\frac{m_i^*}{E_i^*}<p|U_{S,i}(p)+\gamma^{0}U_{V,i}^0(p)|p>.
\end{equation}
The quantities $U_{S,i}$ and $U_{V,i}^{0}$ are momentum dependent,
but to a good approximation can be taken as constant
\cite{Brockmann&Machleidt90,Haar&Malfleit87}. Hence, for the single
particle potential, one can write
\begin{equation}
U_i(p)=\frac{m_i^*}{E_i^*(p)}U_{S,i}+U^{0}_{V,i},
\end{equation}
which, as in Eq.~(12), again amounts to parameterizing the single
particle potential in terms of two constant quantities (for each
type of nucleon). In analogy with the usual Hartree-Fock definition
of the single particle potential (see Eq.~(8)), we also write
\begin{eqnarray}
U_i(p)&=&Re[\sum_{q\leq k_F^n}\frac{m_i^*m_n^*}{E_i^*(p)E_n^*(q)}
<pq|g_{in}|pq-qp>\nonumber\\
&&+\sum_{q\leq k_F^p}\frac{m_i^*m_p^*}{E_i^*(p)E_n^*(q)}
<pq|g_{ip}|pq-qp>]
\end{eqnarray}
where $g_{ij}$ satisfies the relativistic scattering equation
\begin{widetext}
\begin{equation}
g_{ij}({\bf{P,K,K_0}})=\upsilon_{ij}^*({\bf{K,K_0}})-
\int\frac{d^3K'}{(2\pi)^3}\upsilon_{ij}^*({\bf{K,K'}})
\frac{m_i^*m_j^*}{E_i^*E_j^*}
\frac{Q_{ij}({\bf{K',P}};k_F)g_{ij}({\bf{P;K',K_0}})}
{\epsilon_{ij}^*({\bf{P,K'}})-(\epsilon_{ij}^*)_{0}({\bf{P,K_0}})}.
\end{equation}
Using the definitions
$G_{ij}=\frac{m_i^*}{E_i^*}g_{ij}\frac{m_j^*}{E_j^*}$ and
$V_{ij}^*=\frac{m_i^*}{E_i^*}\upsilon_{ij}^*\frac{m_j^*}{E_j^*}$,
Eq. (20) can be rewritten as
\begin{equation}
G_{ij}({\bf{P;K,K_0}})=V_{ij}^*({\bf{K,K_0}})-\int\frac{d^3K'}{(2\pi)^3}
V_{ij}^*({\bf{K,K'}})
\frac{Q_{ij}({\bf{K',P}};k_F)G_{ij}({\bf{P;K',K_0}})}
{\epsilon_{ij}^*({\bf{P,K'}})-(\epsilon_{ij}^*)_0({\bf{P,K_0}})},
\end{equation}
\end{widetext}
which is formally identical to the non-relativistic $G$-matrix
equation (Eq.~(4)). We are then in a situation that is technically
equivalent to the non-relativistic case, except for the Dirac
structure of the single-particle energy. Namely, Eqs.~(19) and (21)
(where the potential $V_{ij}^*$ is written in terms of in-medium
spinors, Eq.~(16)), must be solved self-consistently for $G_{ij}$
and the single particle potential, $U_i(p)$, with the help of the
chosen parametrization, Eq.~(18). We then proceed to calculate the
energy per particle.

In summary, the most significant difference between the BHF and DBHF
schemes is that in the latter case the nucleon wave function is
obtained self-consistently, while in the BHF approach the free-space
solution is used. This difference turns out to be an important one
-- as a result of the reduced nucleon mass in the medium, the lower
component of the nucleon spinor is larger than in free space. This
is well known to produce a density-dependent, repulsive-many body
effect, with the result that the predicted saturation density and
energy are consistent with the empirical values
\cite{Brockmann&Machleidt90}. Physically, the description of the
nucleon in nuclear matter with a Dirac spinor as in Eq.~(16) can be
regarded as effectively taking into account some many-body force
contributions \cite{3B} (hence, the reduced need for inclusion of
three-body forces in the relativistic scheme).

\subsection{Relativistic stellar structure equations}

In the present section we discuss the structure equations of static
(non-rotating) neutron stars. These are objects of highly compressed
matter so that the geometry of spacetime is changed considerably
from that of flat space. Thus models of such stars need to be
constructed within the framework of General Relativity combined with
theories of superdense matter. The connection between these two
branches of physics is provided by Einstein's field equations
\begin{equation}
G^{\mu\nu}=8\pi GT^{\mu\nu}(\epsilon,P(\epsilon)),
\end{equation}
which couple the Einstein curvature tensor, $G^{\mu\nu}$, to the
energy-momentum density tensor, $T^{\mu\nu}$, of stellar matter. The
tensor $T^{\mu\nu}$ contains the equation of state in the form
$P(\epsilon)$ (pressure, $P$, as a function of the energy density,
$\epsilon$), and $G$ is the gravitational constant.

Einstein's field equations are completely general and simple in
appearance. However they are exceedingly complicated to solve
because of their non-linear character and because spacetime and
matter act upon each other. There are only a few cases in which
solutions can be found in closed form. One of the most important
closed-form solutions is the Schwarzschild metric outside a
spherical star. Another is the Kerr metric outside a rotating black
hole.

Starting from Einstein's field equations one can derive the
structure equations of a static, spherically symmetric, relativistic
star. For an explicit treatment and derivation see for example the
work of Fridolin Weber in Ref.~\cite{Weber99}. These equations are
generally known as Tolman-Oppenheimer-Volkoff (TOV) equations. In a
system of units where $c=G=1$, the TOV equations can be written as
\begin{eqnarray}
\frac{dP}{dr}&=&-\frac{\epsilon(r)m(r)}{r^2}
\left[1+\frac{P(r)}{\epsilon(r)}\right]\nonumber\\
&&\times\left[1+\frac{4\pi{r^3}P(r)}{m(r)}\right]
\left[1-\frac{2m(r)}{r}\right]^{-1}
\end{eqnarray}
\begin{equation}
\frac{dm(r)}{dr}=4\pi\epsilon(r)r^{2}dr.
\end{equation}
To proceed to the solution of these equations, it is necessary to
provide the EOS of stellar matter in the form $P(\epsilon)$.
Starting from some central energy density $\epsilon_c=\epsilon(0)$
at the center of the star $(r=0)$, and with the initial condition
$m(0)=0$, the above equations can be integrated outward until the
pressure vanishes, signifying that the stellar edge is reached. Some
care should be taken at $r=0$ since, as seen above, the TOV
equations are singular there. The point $r=R$  where the pressure
vanishes defines the the radius of the star and
$M=m(R)=4\pi\int_0^R\epsilon(r')r'^2dr'$ its gravitational mass.

For a given EOS, there is a unique relationship between the stellar
mass and the central density $\epsilon_c$. Thus, for a particular
EOS, there is a unique sequence of stars parameterized by the
central density (or equivalently the central pressure $P(0)$).

In this work we apply a standard fifth order Runge-Kutta numerical
scheme \cite{NR} to integrate the TOV equations, which are
supplemented by the EOS in numerical form. In the next section, we
discuss the features of the EOS applied in this study.

\section{Equation of state}

We apply here the relativistic equation of state. For a detailed
discussion of the isospin-dependent EOS we refer the reader to
Ref.~\cite{Alonso&Sammarruca1}. In the present study we calculate
neutron star properties considering either pure neutron matter or
asymmetric matter in $\beta$-equilibrium.

\renewcommand{\thesubsection}{3.\arabic{subsection}}
\subsection{Neutron matter EOS}

The degree of isospin asymmetry is represented by the asymmetry
parameter, $\alpha=\frac{\rho_n-\rho_p}{\rho}$ with $\rho_{n,p}$ the
neutron/proton densities and $\rho=\rho_n+\rho_p$ the total density.
(Clearly, $\alpha=0$ for symmetric nuclear matter and $\alpha=1$ for
pure neutron matter.) The isopsin-dependent EOS can be written as
\cite{Alonso&Sammarruca1}
\begin{equation}
\bar{e}(k_F,\alpha)=\frac{(1+\alpha)\bar{e}_n+(1-\alpha)\bar{e}_p}{2}
\end{equation}
with $\bar{e}_{n,p}$ the average energy per neutron/proton and
$\bar{e}(k_F,\alpha)$ the energy per particle as a function of the
total Fermi momentum and the asymmetry parameter. The total Fermi
momentum, $k_F$, is related to the total density, $\rho$, in the
usual way
\begin{equation}
\rho=\frac{2k_F^3}{3\pi^2}.
\end{equation}

We start with showing the isospin-dependent EOS for neutron and
symmetric matter, see Fig.~1. In both cases, predictions are shown
for DBHF and BHF calculations and all three NN potentials referred
to in subsection 2.1. For the case of pure neutron matter (upper
frame of Fig.~1), we see essentially no differences among the
predictions from the three potentials. As already discussed, a main
source of differences among the three versions of the Bonn potential
is in the strength of the tensor force, which is mostly reflected in
the ($T=0$) $^3S_1-^3D_1$ coupled states. In neutron matter ($T=1$),
however, this partial wave does not contribute and thus the model
dependence is dramatically reduced \cite{Li92}. Consistent with
these observations, in symmetric matter significant differences are
observed among the three potentials, see lower panel of Fig.~1.
Notice that our EOSs are considerably more attractive than those
shown in Ref.~\cite{Machleidt89} and also calculated with Bonn A, B,
and C. For details on the differences between our respective
calculations we refer the reader to Ref.~\cite{Alonso&Sammarruca2}.

\begin{figure}[!t]
\centering
\includegraphics[totalheight=3.5in]{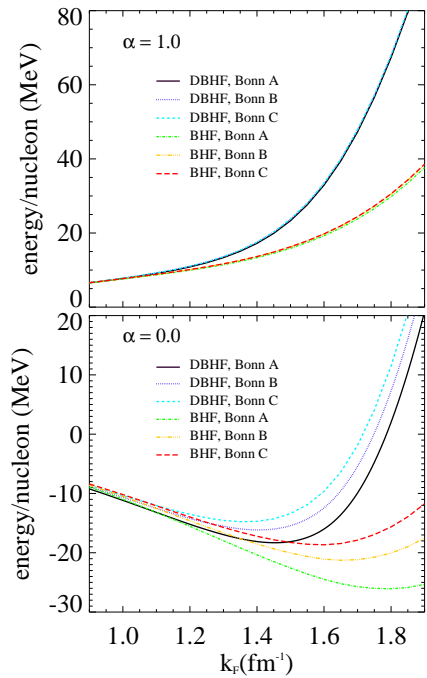}
\caption{(Color online) EOS for neutron ($\alpha=1.0$) and symmetric
matter ($\alpha=0.0$).}
\end{figure}

\begin{figure}[!t]
\centering
\includegraphics[totalheight=2.5in]{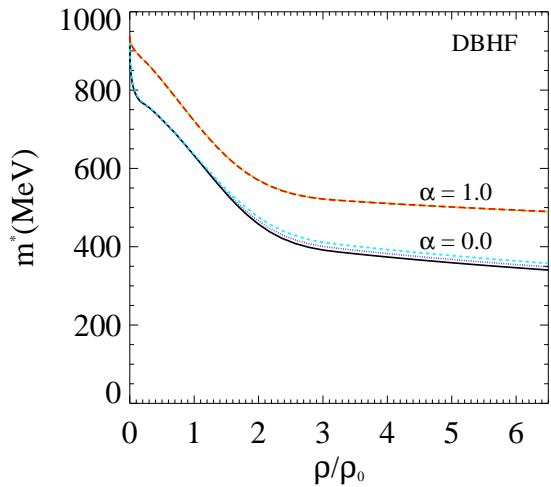}
\caption{(Color online) Effective nucleon masses in neutron
($\alpha=1.0$) and symmetric ($\alpha=0.0$) matter as a function of
density. For $\alpha=0.0$, the solid (black), dotted (dark blue),
and dashed (light blue) curves correspond to Bonn A, B, and C,
respectively. No differences can be seen among the predictions from
the three potentials in the case of neutron matter.}
\end{figure}

For solving the TOV equations one needs to compute energy density
and pressure, both of which are simply related to the EOS. As
previously outlined, in our self-consistent calculation of the EOS
we fit the single-particle potential with its ``effective mass
approximation''. Although chosen on reasonable theoretical grounds,
the {\it ansatz} we use has a finite domain of validity (for which
reliable convergenge of the self-consistent procedure can be
obtained.) On the other hand, for NS calculations one needs to
supply the relationship between energy density and pressure over a
very wide range of densities. The EOS at the lowest/highest
densities have been derived with different methods, usually coupling
available parametrizations of different EOSs
\cite{Ostgaard,Sumiyoshi95}. Here, in the effort of keeping internal
consistency as far as possible, we first obtain nucleon effective
masses by interpolating the available predictions to the free-space
value (to cover the small low-density part, approximately $\rho <
0.03$ fm$^{-3}$) and extrapolating to the higher densities
(approximately $\rho > 0.6$ fm$^{-3}$). With the masses thus
generated, we then proceed to the usual (microscopic) calculation of
the energy per particle everywhere in the needed range. (Notice that
the calculation of the $G$-matrix, and ultimately of the
energy/particle, can proceed as long as effective masses are
provided, as the momentum-independent part of the single-nucleon
potential in Eq.~(18) (or Eq.~(12)) drops out from the energy
denominator in Eq.~(4).)

The effective masses are shown in Fig.~2 for both symmetric and
neutron matter. We note the weaker density dependence at the higher
densities, a behavior that is already reflected in the
self-consistent calculation and appears reasonable on physical
grounds. The nucleon effective masses originate from the
(attractive) scalar potential in the Dirac equation, which, in turn,
is sensitive mostly to the scalar meson $\sigma$. As density
increases, short-range contributions
 become dominant over intermediate-range attraction. This results into
the observed lesser sensitivity of the effective masses to
increasing density.
 (Further discussion on the high-density behavior of EOS
will be presented in subsection 3.3.)

The energy density, $\epsilon$, is defined as
\begin{equation}
\epsilon(\rho,\alpha)=\rho[\bar{e}(\rho,\alpha)+m_N^{free}]
\end{equation}
where $\bar{e}(\alpha,\rho)$ is the energy per nucleon and
$m_N^{free}$ is the nucleon rest mass. The pressure in nuclear
matter is defined in terms of the energy per particle and the baryon
number density as
\begin{equation}
P(\rho,\alpha)=\rho^2\frac{\partial\bar{e}(\rho,\alpha)}{\partial\rho}.
\end{equation}

In the upper frame of Fig.~3 we show the energy density, $\epsilon$,
as a function of the baryon number density, $\rho$, calculated with
the Bonn A, B, and C potentials. The pressure, $P$, as a function of
$\rho$ is shown in the lower panel of Fig.~3.  We also note that our
predictions are reasonably consistent with most recent constraints
on neutron and symmetric matter pressure obtained through medium
energy heavy ion collisions \cite{MSU05}. For the reasons discussed
earlier, hardly any model dependence is seen in the pure $T=1$
system. Accordingly, very small differences  can be expected among
the NS properties computed from pure neutron matter using the three
potentials under consideration. (We reiterate that this is due to
the relationship among these potentials in particular, and does not
imply model independence in general.) Next we discuss the EOS for
$\beta$-stable matter.

\begin{figure}[!t]
\centering
\includegraphics[totalheight=3.5in]{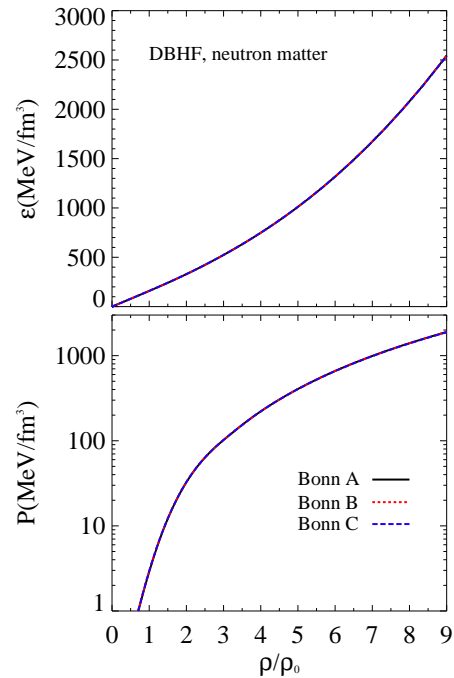}
\caption{(Color online) EOS for pure neutron matter. The upper panel
shows the energy density and the lower panel the pressure as a
function of the baryon number density.}
\end{figure}

\subsection{$\beta$-stable matter EOS}

The density dependence of the symmetry energy determines the proton
fraction in $\beta$-equilibrium, and, in turn, the cooling rate and
neutrino emitting processes.

First, we consider a system of nucleons and electrons only. From the
$\beta$-stability condition, the chemical potential of the electrons
is written as
\begin{equation}
\mu_n-\mu_p=\mu_e
\end{equation}
Deriving the EOS of $\beta$-stable matter requires knowing the
corresponding proton fraction, $Y_p=\frac{\rho_p}{\rho}$. The
equilibrium particle concentrations,  $Y_i=\frac{\rho_i}{\rho}$
($i=e$, $n$, $p$), can be calculated via Eq.~(29) combined with the
charge neutrality condition $\rho_e=\rho_p$ (i.e. $Y_e=Y_p$) and
$\rho=\rho_n+\rho_p$ (i.e. $1=Y_n+Y_p$). To compute the various
chemical potentials,
\begin{equation}
\mu_i=\frac{\partial e_{tot}}{\partial Y_i},
\end{equation}
we write  $e_{tot}$ as
\begin{equation}
e_{tot}(\rho,Y_p,Y_n,Y_e)=\bar{e}(\rho,Y_p)+\frac{\rho_p}{\rho}m_pc^2+
\frac{\rho_n}{\rho}m_nc^2+\frac{E_e}{A}.
\end{equation}
Here $\bar{e}(\rho,Y_p)$ is the energy per nucleon and
$\frac{E_e}{A}$ is the contribution to the total energy from the
electrons. Similarly to what is done in
Ref.~\cite{Bombaci&Lombardo91}, we assume a simple model for
electrons and treat them as a gas of extremely relativistic
non-interacting fermions. That is, for electrons we set
\begin{equation}
c(p^2+m_e^2c^2)^{1/2} \approx pc,
\end{equation}
which appears reasonable since the electron mass $m_e$ is small
compared to its chemical potential at typical nuclear density.
Integrating over all momentum states (and summing over spin states),
it is easy to show that the electron energy density (or energy per
volume) becomes
\begin{equation}
\frac{E_e}{V} = \frac{(k_F^e)^4}{4 \pi ^2} =
\frac{(3\pi^2{\rho}Y_e)^{4/3}}{4\pi^2}
\end{equation}
where $k_F^e$ is the electron Fermi momentum.

In Ref.~\cite{Alonso&Sammarruca1} we have verified that the energy
per nucleon can be written as
\begin{equation}
\bar{e}(k_F,\alpha)=\bar{e}(k_F,0)+e_{sym}(k_F)\alpha^2,
\end{equation}
with $\bar{e}(k_F,0)$ the energy per particle for symmetric matter
and $e_{sym}$ the symmetry energy defined by
\begin{equation}
e_{sym}(k_F)=\left.\frac{1}{2}
\frac{\partial^2\bar{e}(k_F,\alpha)}{\partial^2\alpha}\
\right|_{\alpha=0}.
\end{equation}
With the definition $\alpha=1-2Y_p$, Eq.~(34) becomes
\begin{equation}
\bar{e}(\rho,Y_p)=\bar{e}(\rho,Y_p=\frac{1}{2})+e_{sym}(\rho)(1-2Y_p)^2.
\end{equation}
Thus, $e_{tot}$ is written as
\begin{eqnarray}
&&e_{tot}(\rho,Y_p,Y_n,Y_e)=\bar{e}(\rho,Y_p=\frac{1}{2})+e_{sym}(\rho)(1-2Y_p)^2\nonumber\\
&&+ Y_p m_pc^2 + Y_n m_nc^2+\hbar c
\frac{(3\pi^2Y_e)^{4/3}}{4\pi^2}\rho^{1/3},
\end{eqnarray}
where the last therm can be obtained trivially from Eq.~(33) divided
by the density. Knowing the total energy allows one to evaluate the
chemical potentials, which, in turn, are inserted in Eq.~(29)
(combined with the constraint $Y_p=Y_e$) to give a simple algebraic
equation for $Y_p$
\begin{equation}
-4e_{sym}(\rho)(1-2Y_p)+(m_p-m_n)c^2+
\hbar{c}(3\pi^2\rho)^{1/3}Y_p^{1/3}=0.
\end{equation}
The solution of this equation is inserted in Eq.~(36) to provide the
EOS of nuclear matter in $\beta$-equilibrium with electrons.

Just above the density of normal nuclear matter, the electron
chemical potential, $\mu_e$, exceeds the muon mass, $m_{\mu}$, and
the reaction $n\leftrightarrow p+\mu^{-}$ becomes energetically
allowed~\cite{WFF88}. Under these circumstances the
$\beta$-stability condition and charge-neutrality require
\begin{equation}
\mu_e=\mu_{\mu},
\end{equation}
and
\begin{equation}
\rho_p=\rho_e+\rho_{\mu}
\end{equation}
(and thus, $Y_p-Y_e-Y_{\mu}=0$). Equation~(31) needs to be modified
to include the muon contribution to the total energy,
\begin{eqnarray}
e_{tot}(\rho,Y_p,Y_n,Y_e,Y_{\mu})&=&\bar{e}(\rho,Y_p)+Y_pm_pc^2+Y_nm_nc^2\nonumber\\
&+&\frac{E_e}{A}+\frac{E_{\mu}}{A}.
\end{eqnarray}

We will treat muons as a gas of non-interacting nonrelativistic
fermions \cite{MBBS04}, a reasonable approximation given the
relatively large muon mass (compared to its Fermi momentum at
typical muon concentrations in $\beta$-equilibrium). It can then be
shown that the muon energy density is~\cite{Glendenning}
\begin{equation}
\frac{E_{\mu}}{V}\approx
\rho_{\mu}m_{\mu}+\frac{(3\pi^2\rho_{\mu})^{5/3}}{10\pi^2m_{\mu}}
\end{equation}
(in units where $c=\hbar=1$). Thus, Eq.~(41) becomes
\begin{eqnarray}
&&e_{tot}(\rho,Y_p,Y_n,Y_e,Y_{\mu})=\bar{e}(\rho,Y_p=\frac{1}{2})+e_{sym}(\rho)(1-2Y_p)^2\nonumber\\
&&+ Y_p m_pc^2 + Y_n m_nc^2+{\hbar
c}\frac{(3\pi^2Y_e)^{4/3}}{4\pi^2}\rho^{1/3}\nonumber\\
&&+Y_{\mu}m_{\mu}c^2+\frac{(\hbar
c)^2}{2m_{\mu}c^2}\frac{(3\pi^2Y_{\mu})^{5/3}}{5\pi^2}\rho^{2/3},
\end{eqnarray}
where the muon contribution to the total energy per particle is
obtained from Eq.~(42) divided by $\rho$. As done in case of
nucleons and electrons only, evaluating the chemical potentials,
combined with Eqs.~(39) and (40) leads to the following algebraic
equations for the equilibrium particle concentrations ($Y_i$)
\begin{equation}
-4e_{sym}(\rho)(1-2Y_p)+(m_p-m_n)c^2+
\hbar{c}(3\pi^2\rho)^{1/3}Y_e^{1/3}=0
\end{equation}
\begin{equation}
m_{\mu}c^2+({\hbar}c)^2\frac{(3\pi^2)^{5/3}}{6\pi^2m_{\mu}c^2}\rho^{2/3}Y_{\mu}^{2/3}
-\hbar{c}(3\pi^2\rho)^{1/3}Y_e^{1/3}=0
\end{equation}
\begin{equation}
Y_p-Y_e-Y_{\mu}=0.
\end{equation}
Once $Y_p$ is obtained from the above equations, it is inserted in
Eq.~(36) to provide again the EOS of $\beta$-stable matter. The
energy density of baryon/lepton matter in $\beta$-equilibrium is
then used as the new input for the TOV equations.

\begin{figure}[!t]
\centering
\includegraphics[totalheight=3.5in]{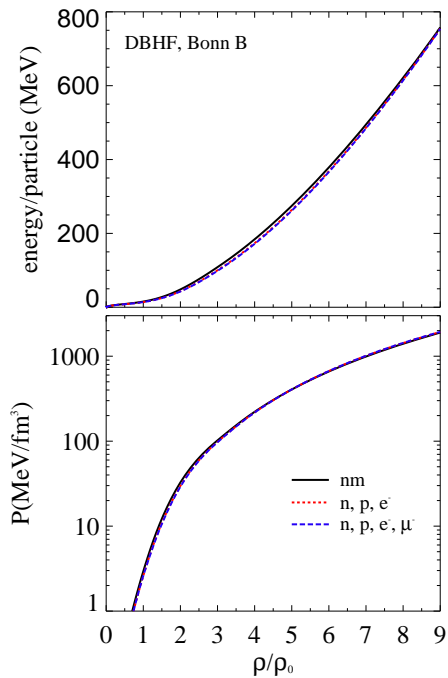}
\caption{(Color online) EOS for neutron matter and baryon/lepton
matter in $\beta$ equilibrium. Predictions are obtained with the
Bonn B potential. The upper panel shows the energy density and the
lower panel the pressure as a function of the baryon number
density.}
\end{figure}

In Fig.~4 we compare EOSs for neutron matter and $\beta$-stable
matter (for both $e^-$ and $e^-$ plus $\mu^-$ cases), obtained with
the Bonn B potential. We observe only very small differences among
the various predictions, although the $\beta$-stable EOSs are
slightly ``softer'' than the one for pure neutron matter (at
intermediate densities), a trend that is independent of the
particular NN potential being used. This is to be expected,  since a
system with some $T$=0 component is generally more attractive
(mostly through the $^3S_1$ partial wave) than a pure $T$=1 system.
This point will be looked at more closely in the next subsection.

The symmetry energy and the proton fraction in $\beta$-stable matter
are displayed in Fig.~5. We observe that the symmetry energy (upper
frame) tends to saturate and eventually decreases with density, a
behavior qualitatively similar to the one seen in the proton
fraction (middle frame). From the lower panel of Fig.~5, we see that
the addition of muons increases the proton fraction noticeably. In
particular, the maximum proton fraction increases from 0.11 to about
0.13 (with Bonn B). We recall that a proton fraction larger than
approximately $\frac{1}{9}$ would allow a critical cooling mechanism
for the neutron star via the direct URCA processes \cite{LP2004}. On
the other hand, due to the behavior of the predicted symmetry energy
(the crucial player in the determination of the proton fraction),
our proton concentrations remain very small, and thus the EOS is
only mildly impacted by the inclusion of leptons. In Fig.~6 we show
the different particle concentrations as a function of density. The
presence of muons becomes possible only at higher densities, due to
the energetic constraints mentioned earlier. Also, their
concentration remains substantially below the electron fraction.

In closing this section, we stress that the high-density behavior of
the symmetry energy is very poorly constrained and theoretically
controversial, with different models often yielding dramatically
different predictions. A more detailed discussion on this and
related issues is presented next.

\begin{figure}[!t]
\centering
\includegraphics[totalheight=4.0in]{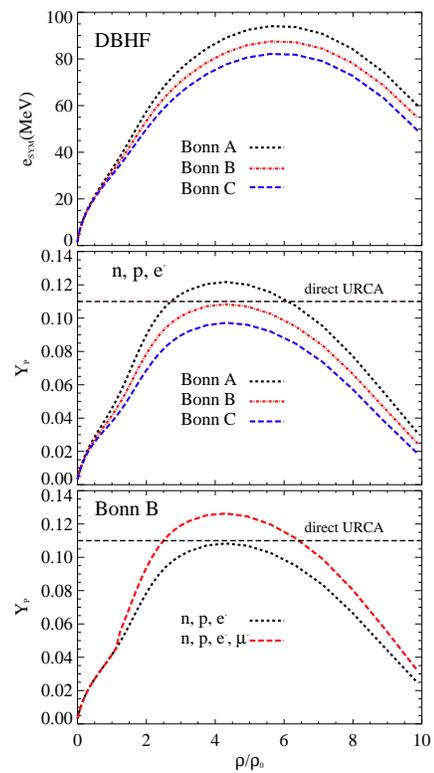}
\caption{(Color online) Symmetry energy, $e_{sym}$, and proton
fraction in $\beta$-equilibrium, $Y_P$. The upper panel shows the
symmetry energy and the middle/lower panel the proton fraction as a
function of density.}
\end{figure}

\begin{figure}[!t]
\centering
\includegraphics[totalheight=2.25in]{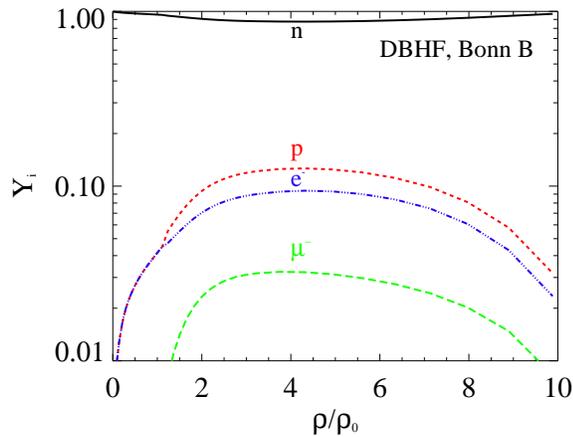}
\caption{(Color online) Concentrations of various particle species
($Y_i=\rho_i/\rho$) in $\beta$-stable matter  as a function of the
baryon number density in units of $\rho_0$.}
\end{figure}

\subsection{High-density equation of state and the meson model:
further discussion}

In this subsection, we take a more in-depth look at how the observed
features of the EOS and the symmetry energy can be understood in
relation to the physical components of our model.

First, we observe that from Eq.~(34) one can write
\begin{equation}
\bar{e}(k_F,\alpha=1)-\bar{e}(k_F,\alpha=0)=e_{sym}(k_F),
\end{equation}
which clearly displays the significance of the symmetry energy as
the energy shift between neutron and symmetric matter. Thus, the
behavior of $e_{sym}$ reflects the relationship between
$\bar{e}(k_F,\alpha=1)$ and $\bar{e}(k_F,\alpha=0)$ as a function of
density. What we observe in the present calculation is that the
energy per particle in symmetric matter becomes more and more
repulsive with increasing density, eventually approaching the one of
neutron matter. Within the meson model, this can be understood in
terms of the competing roles of the intermediate-range attraction
(provided by the scalar meson $\sigma$ and the iterated one-pion
exchange) and the short-range repulsion (generated by the vector
meson $\omega$), as we describe next. Neutron matter is generally a
more repulsive system because it lacks the contribution from T=0
waves, some of which are very attractive at normal nuclear densities
(hence, for instance, the crucial role of the $^3S_1$ state in
nuclear binding). On the other hand, as density increases,
short-range repulsion becomes the dominant contribution to the NN
interaction. Furthermore, in the DBHF calculations the (repulsive)
spin-orbit interaction is strongly enhanced due to the relativistic
effective mass in the OBE potential. Therefore, keeping in mind that
only T=1 states contribute to the energy of neutron matter while
both isospin states contribute to the energy of nuclear matter, if
major T=0 partial waves become increasingly repulsive at short
distances, it is possible for the energy of symmetric matter to grow
at a faster rate and eventually approach the neutron matter EOS.
This is just what we observe in our model, as can be seen from
Fig.~7, where we compare the energy per particle in symmetric and
neutron matter up to very high densities. (The arguments above also
explain why the EOSs for neutron matter and $\beta$-stable matter
are very close to each other, cf.~Fig.~4.)

\begin{figure}[!t]
\centering
\includegraphics[totalheight=2.25in]{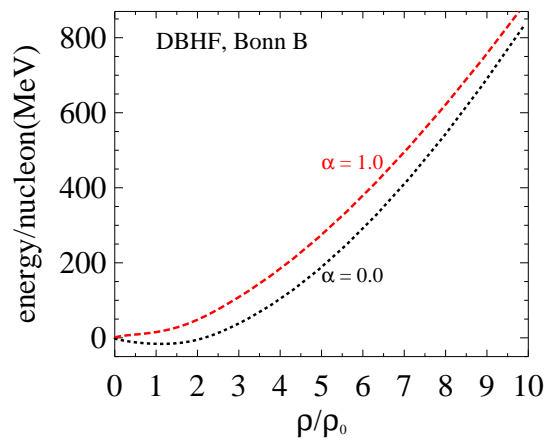}
\caption{(Color online) Energy per nucleon for neutron (red curve)
and symmetric (black curve) matter.}
\end{figure}

To explore this further as well as check internal consistency, we
have performed some diagnostic tests. We find that the contributions
to the average potential energy of symmetric matter from {\it only}
$^1S_0$ are -18.75 MeV, -22.75 MeV, and 5.554 MeV at $k_F$=1.4, 2.0,
and 2.5 fm$^{-1}$, respectively. Similarly, including {\it only} the
$^3S_1$ and $^3D_1$ states, we find (for the same Fermi momenta)
-18.74 MeV, -10.04 MeV, and 18.16 MeV. A correct way to state the
result of the above tests is to say that, in the presence of
repulsive forces only (a scenario simulated by the presence of
high-density), nuclear matter would be a more repulsive system than
neutron matter (for the same $k_F$).

At some critical density, signaled by the symmetry energy turning
negative, it would then be possible for a pure neutron system to
become more stable than symmetric matter, a phenomenon referred to
as {\it isospin separation instability} \cite{Li02}. Clearly, the
value of such critical density depends upon the relative degrees of
attraction and repulsion in the particular model under
consideration. (Figure 5 suggests that in the present model this
would happen at densities well above ten times nuclear matter
density.)

It is of course possible that contributions not included in our
model would soften the EOS in such a way as to alter the balance
between the curves shown in Fig.~7 and, in turn, the high-density
behavior of the symmetry energy. This is precisely among the aspects
we wish to learn about with this and future studies (namely, domain
of validity of the meson-theoretic picture, density dependence of
the repulsive core, etc...). Systematic calculations based not on
phenomenology but on a consistent theoretical framework (in spite of
its inherent limitations), together with stringent constraints, can
help achieve that purpose. Our main conclusion at this point is that
empirical constraints specifically on the high-density behavior of
the symmetry energy would provide some clear and direct information
on the short-range nature of the nuclear force.

\section{Results for neutron star total masses and radii}

In Fig.~8 we show the total masses and radii versus the stellar
central density, $\rho_c$. As expected in the light of our previous
discussion, the NS properties computed from the EOS of neutron
matter are essentially the same for all of the three potentials
applied here. Consistent with the observations contained in
subsection 3.3, the $\beta$-stable matter EOS, being only slightly
softer than the pure neutron matter EOS, yields similar predictions,
although radii tend to be smaller at higher central densities. In
summary the neutron matter EOS from DBHF calculations yields a
maximum mass $M_{max}\simeq 2.25$ $M_{\odot}$ at radius $R\simeq
11.01$ km and central density $\rho_c\simeq 0.98$ $fm^{-3}$. The
relativistic EOS for $\beta$-stable matter yields maximum masses in
the range $M_{max}\approx (2.238 - 2.241)$ $ M_{\odot}$ at radii
$R\approx (10.74 -10.87)$ km and central densities $\rho_c\approx
(0.998 - 1.013)$ $fm^{-3}$. Including muons in the description of
$\beta$-stable matter does not alter significantly the predicted NS
properties from the electrons-only case. Table 1 provides an
overview of our predictions. Figure 9 displays the mass-radius
relation for the NS models shown in Fig.~8.

Before we proceed with our discussion, we recall (see section 3)
that an extrapolation is applied to the nucleon effective mass at
those densities where a self-consistent solution for the
single-particle potential is not available. Of course, every
numerical extrapolation/interpolation involves uncertainties of some
degree. To investigate to what extent the NS properties are
sensitive to variations of $m^*$, we have performed numerous tests
all of which verify that the predicted NS masses and radii are not
affected significantly by moderate variations of the nucleon
effective mass. For instance, uncertainty of 20\% in the high
density effective mass results in approximately 2.9\% variation in
the NS maximum mass, 1.54\% in the radius, and about 0.31\% in the
central density (e.g. Bonn B, DBHF, neutron matter:
$M_{max}=2.2457\pm 0.0659M_{\odot}$, $R=11.001\pm 0.17$ km,
$\rho_c=0.979\pm 0.003fm^{-3}$). Very similar conclusions apply for
the case of $\beta$-stable matter. (Sensitivity tests are performed
in the following way: the highest density mass in Fig.~2 is varied
by up to 20\% and the self-consistently calculated values are then
joined to it by interpolation. In this way the applied mass
variation increases gradually as we move to higher densities, where
the uncertainty should be largest.)

\begin{figure}[!t]
\centering
\includegraphics[totalheight=4.3in]{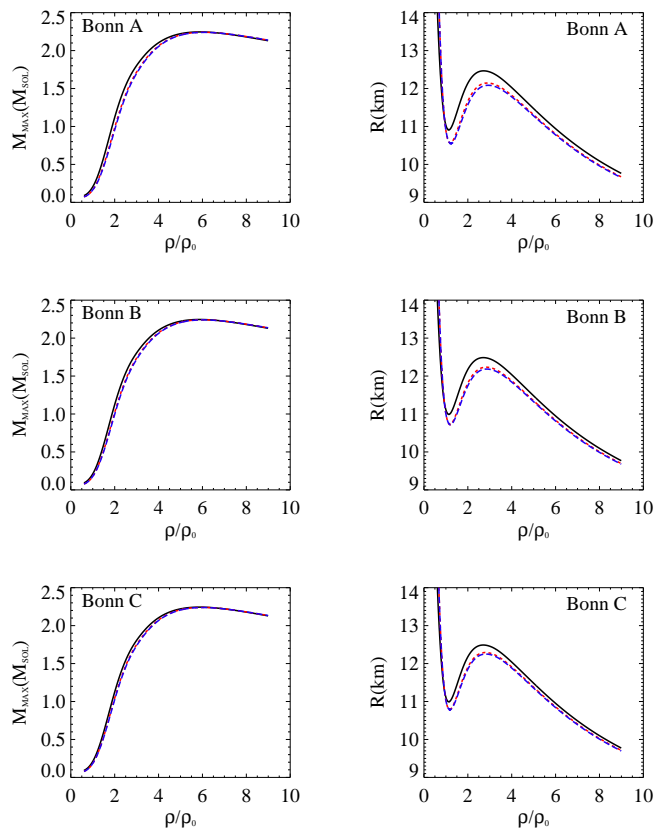}
\caption{(Color online) NS total masses and radii. In all frames the
solid (black) curve corresponds to the DBHF calculation for pure
neutron matter, the short-dashed (red) curve to the DBHF calculation
for $\beta$-stable matter of nucleons and electrons only, and the
long-dashed (blue) curve to the DBHF calculation for nucleons,
electrons and muons in $\beta$-equilibrium.}
\end{figure}

\begin{figure}[!h]
\centering
\includegraphics[totalheight=1.7in]{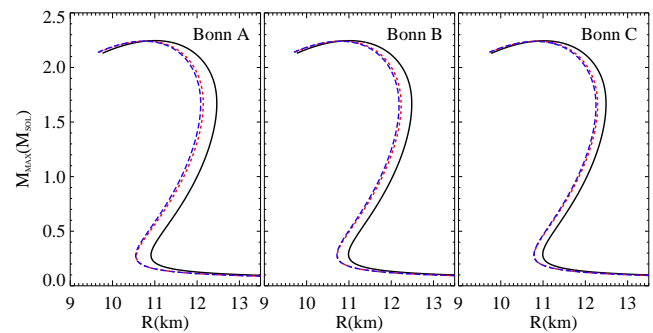}
\caption{(Color online) Mass-radius relation. The curves are labeled
as in Fig.~8.}
\end{figure}

\renewcommand{\thetable}{\arabic{table}}
\begin{table}[!b]
\caption{NS maximum masses, radii, and central densities (from DBHF
calculations).}\vspace{1mm}

\centering
\begin{tabular}{ccccc}
potential model & composition & $M_{max}(M_{\odot})$ & $R(km)$
& $\rho_c(fm^{-3})$\\
\hline\hline
Bonn A & n                   & 2.2456    & 11.00  &  0.979\\
Bonn B & n                   & 2.2453    & 11.01  &  0.979\\
Bonn C & n                   & 2.2439    & 11.02  &  0.978\\
\hline
Bonn A & n,p,$e^{-}$         & 2.2414    & 10.78  &  1.007\\
Bonn B & n,p,$e^{-}$         & 2.2413    & 10.83  &  1.002\\
Bonn C & n,p,$e^{-}$         & 2.2399    & 10.87  &  0.998\\
\hline
Bonn A & n,p,$e^{-},\mu^-$   & 2.2401    & 10.74  &  1.013\\
Bonn B & n,p,$e^{-},\mu^-$   & 2.2399    & 10.79  &  1.008\\
Bonn C & n,p,$e^{-},\mu^-$   & 2.2384    & 10.83  &  1.003\\
\hline\hline
\end{tabular}
\end{table}

Our findings are generally consistent with previous ones obtained
from similar theoretical frameworks
\cite{Ostgaard,Baldo&Burgio98,Sumiyoshi95}. In Ref.~\cite{Ostgaard},
for instance, both BHF and DBHF approaches are used together with
the Bonn A potential. Using the relativistic EOS, the authors
predict a maximum mass $M_{max}\approx 2.37$ $M_{\odot}$ with a
radius $R\approx 12.2$ km at a central density of $\rho_c\approx
0.8$ $fm^{-3}$. We must keep in mind that the radius of a neutron
star is mostly sensitive to the difference in pressure between
neutron and symmetric matter \cite{Li&Steiner05} (the same mechanism
that pushes neutrons out in the skin of a large nucleus). We also
recall that, although our EOS's are rather repulsive (a feature
generally shared by DBHF predictions), for the reasons discussed in
subsection 3.3 the symmetry energy does not keep on growing with
increasing density. As a consequence, the radii we predict are
smaller than, for instance, those reported in Ref.~\cite{Ostgaard}.
From Fig.~9, we observe that our predictions are compatible with
recent findings from analysis of heavy-ion collision observables
\cite{Li&Steiner05} which constrain the radius of a $1.4$
$M_{\odot}$ neutron star to be between $11.5$ and $13.6$ km.

The ultimate upper limit of the NS mass can be deduced theoretically
on the basis \cite{Rhoades&Ruffini74} that (1) general relativity is
the correct theory of gravity, (2) the EOS of stellar matter is
known below some matching density, and (3) the EOS of NS matter
satisfies both (i) the causality condition $\frac{\partial
P}{\partial\epsilon}\le c^2$ and (ii) the microscopic stability
condition $\frac{\partial P}{\partial\epsilon}\ge 0$ (known as the
Le Chatelie's principle). Under the above assumptions the upper mass
limit was found to be $3.2$ $M_{\odot}$. It is generally accepted
that stars with masses above $3.2$ $M_{\odot}$ collapse to black
holes. On the other hand, there is a practical theoretical lower
limit for the NS gravitational mass. The minimum possible mass is
evaluated to be about $(1.1-1.2)$ $M_{\odot}$ and follows from the
minimum mass of a proto-neutron star. This is estimated by examining
a lepton-rich configuration with a low-entropy inner core of
approximately $0.6$~$M_{\odot}$ and a high-entropy envelope
\cite{Goussard98}. This argument is in general agreement with the
theoretical results of supernova calculations in which the inner
collapsing core material comprises at least $1.0$ $M_{\odot}$
\cite{Lattimer2001}.

On the experimental side, the masses of neutron stars are mainly
deduced by observations of NS binary systems.  Precise measurements
of the masses of the binary pulsar PSR 1936+16 yield $1.344$
$M_{\odot}$ and $1.444\pm 0.008$ $M_{\odot}$
\cite{Weisberg&Taylor84}. Recently a few other binary pulsars have
been observed all having masses in the narrow range $1.35\pm 0.04$
$M_{\odot}$ \cite{Thorsett&Chakrabarty99}. In addition, there are
several X-ray binary masses that have been measured to be well above
the average of $1.4$ $M_{\odot}$. The non-relativistic pulsar PSR
J1012+5307 is believed to have a mass of approximately $2.35\pm
0.85$ $M_{\odot}$ \cite{vanKerkwijk:1996sx}. Also the mass of the
Vela X-1 pulsar has been deduced to be approximately $1.9\pm 0.2$
$M_{\odot}$ \cite{Quaintrell:2003pn}. Another object of the same
type is Cygnus X-2 with a mass of $1.78\pm 0.23$ $M_{\odot}$
\cite{King:1998vp}. More recently, the mass of the binary
millisecond pulsar PSR J0751+1807 has been measured (by relativistic
orbital decay) to be $2.1\pm 0.2 M_{\odot}$~\cite{Nice2005}, which
makes this neutron star the most massive ever detected.

Although accurate masses of several neutron stars are available,
precise measurements of the radii are not yet available. (As
mentioned earlier, terrestrial nuclear laboratory data are presently
being used to obtain constraints on NS radii.) It has been shown
that the causality condition can be used \cite{Lattimer90} to set
the lower limit of the radius to about $4.5$ km. In general,
estimates of NS radii from observations have given a wide range of
results. Perhaps the most reliable estimates can be done on the
basis of thermal emission observations from neutron stars surfaces
which yield values of the so-called radiation radius
\begin{equation}
R_{\infty}=\frac{R}{\left(1-\frac{2GM}{Rc^2}\right)^{\frac{1}{2}}}
\end{equation}
a quantity related to the red-shift of the star's luminosity and
temperature \cite{Lattimer2001}. The authors of
Ref.~\cite{Golden&Shearer1999} give values of $R_{\infty}\leq 9.5$
km for a black-body source and $R\leq 10.0$ km for an object with a
magnetized H atmosphere. The discovery of the quasi-periodic
oscillations from X-ray emitting neutron stars in binary systems
provides a possible way of constraining the NS masses and radii.
Einstein's general relativity predicts the existence of a maximum
orbital frequency which yields a mass of approximately $1.78$
$M_{\odot}$ at a radius of about $8.86$ km. Including corrections
due to stellar rotation induces small changes in these numbers
\cite{Psaltis1998}. Recently, a radio pulsar spinning at 716 Hz, the
fastest spinning neutron star to date, has been
discovered~\cite{Hessels:2006ze}. The authors of
Ref.~\cite{Hessels:2006ze} conclude that if the pulsar has a mass
less than 2 $M_{\odot}$, then its radius is constrained by its spin
rate to be $<16$ km. Finally, the latest observations from EXO
0748-676 are reported to be consistent with the constraints $2.10\pm
0.28$ $M_{\odot}$ and $13.8\pm 1.8$ km for the mass and the radius,
which favor relatively stiff equations of state \cite{nature06}.

In view of the above survey of presently available experimental
constraints and/or estimates, we may conclude that our relativistic
predictions are reasonable, although our values for the maximum
masses are on the high side of the presently accepted range. On the
other hand, it must be kept in mind that mechanisms not included in
the present calculation but generally agreed to take place in the
star interior would further soften the EOS, thus decreasing the
computed NS mass and radius. Among these mechanisms are pion and
kaon condensations, increase in the hyperon population due to rising
chemical potentials with density, and transition to quark matter.
All these phenomena tend to lower the energy per particle, and
decrease the NS mass and radius while rising the stellar central
density. Considering the effect of rotation in the NS properties
calculation would increase the mass (by about 15\%) and so this
repulsive contribution may be in part ``cancelled'' by any of the
attractive mechanisms mentioned above. In summary, it seems that
inclusion of additional degrees of freedom would, overall, move our
DBHF predictions in the right direction. Our conclusions are
summarized in the next section.

\section{Conclusions}

We  have presented  systematic calculations of NS limiting masses
and radii using relativistic EOSs. We have considered the case of
pure neutron matter and nuclear matter in $\beta$-equilibrium. The
NS properties obtained in each case are very similar to each other,
a behavior closely related to the predicted density dependence of
the symmetry energy and the proton fraction in chemical equilibrium.
The present analysis helped us establish a clear correlation between
the high-density behavior of the symmetry energy and the nature of
the repulsive core as it manifests itself in specific partial waves.
We have discussed this issue in considerable details and stressed
the importance of reliable experimental information to set stringent
constraints for theoretical predictions of the symmetry energy at
high density.

After overviewing presently available empirical information, we
conclude that our DBHF-based predictions of NS maximum masses are
somewhat high, indicating that repulsion in the high-density EOS
should be reduced (again, keeping in mind that the difference in
pressure between neutron and symmetric matter will primarily control
the radii).

It is likely that other mechanisms, not included in the present
hadronic model, may take place in the stellar interior which are
likely to soften the EOS. Pions and kaons may be likely to
condensate in the interior of neutron stars
~\cite{Takahashi:2002ig,Muto:2005my,Kubis:2002dr}. It is generally
agreed upon that pion/kaon populations increase the proton fraction
and might cause a rapid cooling via the direct URCA process. The
stellar matter EOS could soften considerably due to pion/kaon
condensations, since these mechanisms tend to lower the symmetry
energy. In addition, at higher densities different species of
hyperons appear in the NS composition which also results in a
softening of the stellar matter EOS. The expected transition to
quark matter at higher densities would lower the energy and soften
EOS. Therefore, the present set of results give us confidence in our
relativistic microscopic approach as a reasonable baseline from
which to proceed.

Another important issue is the effect magnetization may have on the
EOS, namely, how the energy per particle changes if matter in the
stellar interior becomes spin-polarized. As a next step towards a
better and more complete understanding of the physics of neutron
stars, we are presently extending our framework to include the
description of spin-polarized neutron/nuclear matter.

Finally, the observations of compact stars will be greatly improved
in the future by the Square Kilometer Array (SKA). The SKA is an
internationally sponsored project with the goal to construct a
radio-telescope with a total receiving surface of one million square
meters. The SKA is a facility with a potential to detect from 10,000
to 20,000 new pulsars, more than 1,000 millisecond pulsars and at
least 100 compact relativistic binaries \cite{Kramer:2003xs}. These
future endeavors will provide a tool capable of probing the NS
matter EOS at the extreme limits.

\section*{Acknowledgements}

\noindent The authors acknowledge financial support from the U.S.
Department of Energy under grant number DE-FG02-03ER41270.

\end{document}